# Heavy ion acceleration in the Breakout Afterburner regime

G. M. Petrov[1], C. McGuffey[2], A. G. R. Thomas[3] and K. Krushelnick[3] and F. N. Beg[2]

[1]Naval Research Laboratory, Plasma Physics Division, 4555 Overlook Ave. SW, Washington, DC 20375, USA

[2]Mechanical and Aerospace Engineering and Center for Energy Research, University of California-San Diego, La Jolla, CA, 92093 USA

[3]Center for Ultrafast Optical Science, University of Michigan, Ann Arbor, MI 48109 USA

e-mail: george.petrov@nrl.navy.mil

## Abstract

Theoretical study of heavy ion acceleration from an ultrathin (20 nm) gold foil irradiated by sub-picosecond lasers is presented. Using two dimensional particle-in-cell simulations we identified two highly efficient ion acceleration schemes. By varying the laser pulse duration we observed a transition from Radiation Pressure Acceleration to the Breakout Afterburner regime akin to light ions. The underlying physics and ion acceleration regimes are similar to that of light ions, however, nuances of the acceleration process make the acceleration of heavy ions more challenging. Two laser systems are studied in detail: the Texas Petawatt Laser and the Trident laser, the former having pulse duration 180 fs, intermediate between very short femtosecond pulses and picosecond pulses. Both laser systems generated directional gold ions beams (~10 degrees half-angle) with fluxes in excess of $10^{11}$ ion/sr and normalized energy >10 MeV/nucleon.

# 1. Introduction

Ion beams generated from high-Z material are useful for many applications [1,2,3] including nuclear reactions [4,5], production of super-heavy elements [6], exotic isomers and isotopes for biomedical use, fast ignition fusion, radiation effects in materials [7], medical applications including radiotherapy [8] and radiation oncology [9] and studies of exotic phenomena such as systems relevant to the interior of stars. So far, research has been conducted using conventional accelerators at enormously large and expensive facilities, however, short pulse lasers with duration <1 ps are emerging as a viable alternative tool for their production, making possible laboratory experiments at much lower cost. Laser-driven accelerators can generate heavy ion beams with energy in excess of 100s of MeV. The beams are highly directional, which results in relatively large fluxes, on the order of $10^{11}$ $ions/sr$. In spite of the potential short pulse lasers offer, this approach is still largely unexplored, though possible applications of laser-driven heavy ion beams have been discussed in a recent paper by Nishiuchi *et al*. [10].

In our previous publication, we studied theoretically the generation of heavy ion beams driven by a short pulse laser with duration 32 fs using 2D particle-in-cell (PIC) simulations [11]. These are one of the shortest pulses available today at intensities relevant to our study ($I_0 > 10^{20} W/cm^2$). We established that the acceleration of gold ions is due to either Target Normal Sheath Acceleration (TNSA) [12] or Radiation Pressure Acceleration (RPA) [13,14,15,16,17]. Which acceleration mechanism dominates depends primarily on the foil thickness. For very thin foils, <100 nm, in the RPA regime the simulations predicted gold ions with energy ~10 MeV/nucleon can be generated with conversion efficiency (into gold ions) of ~8 %. The theoretical study revealed that for short pulses the conditions are far more demanding compared to protons and light ions and a number of challenges must be overcome in order to successfully accelerate heavy ions. Among them is the limited charge-to-mass ratio *q/M*, resulting in lower energy per nucleon compared to light ions due to the unfavorable scaling of the normalized ion energy with *q/M*: $E/M \sim (q/M)^2$ [18,19]. In particular, numerical simulations established that for gold $q/M < 0.3$. Due to the low *q/M* it takes longer to accelerate the ions. Finally, it was established that only one half of the laser pulse can be used for acceleration, since the ions must be first ionized, which occurs near the peak of the pulse, before being accelerated. We concluded that acceleration of gold ions using very short (30-50 fs) pulses is borderline possible. The difficulties accompanying the acceleration (low *q/M* and short acceleration time) can, in principle, be compensated by increasing the laser intensity, which, in turn, may invoke other issues.

The alternative is to use longer pulses. A "long" (ps) pulse would provide ample acceleration time, which eliminates the hurdles discussed above. It is now widely recognized and proved experimentally that they work well for protons and light ions in the so called Breakout Afterburner (BoA) [20,21,22,23] regime. The next logical step is to address the more challenging case of heavy ions such as Au or W. To our knowledge, there is only one experimental work using very short (~40 fs) pulses [24] and another using long (~1 ps) pulses [25] and no theoretical studies in the BoA regime. The purpose of this work is to extend the study of heavy ion acceleration from femtosecond to picosecond pulses and to explore other regimes of ion acceleration, more specifically, BoA. It is also of fundamental interest to understand the transition from RPA to BoA for heavy ions and for this purpose we consider modeling a laser system with pulse duration intermediate between femtosecond and picosecond. An appropriate system with this property is the Texas Petawatt Laser, which has pulse duration of 180 fs. In Section 2 we will briefly discuss the RPA-to-BoA transition by carrying out 2D3V PIC numerical simulations for gold ion acceleration from sub-micron foils for three pulse durations. In Section 3 we discuss in more detail numerical results for the Texas Petawatt Laser, while in Section 4 simulation results for the Trident laser are presented. A discussion and summary are given in the final sections of the paper.

## 2. Transition from RPA to BoA

Numerical simulations are performed using a two-dimensional electromagnetic PIC code [26,27]. In all cases the target is a flat 20 nm Au foil covered with a 5 nm contaminant layer residing on the back of the foil, located at spatial position $x = 48\,\mu m$. For numerical purposes, the contaminants are modeled as a thin sheet of water at liquid density. The foil thickness is chosen to roughly correspond to the relativistic skin depth $\ell_{skin} = \gamma^{1/2} c/\omega_p$ at the peak of the laser pulse, where $\gamma$ and $\omega_p$ are the relativistic parameter and electron plasma frequency, and $c$ is the speed of light. Under these conditions, the laser field can penetrate the whole target and volumetrically heat the electrons in the entire volume in the laser spot. This is essential for the initial stage of BoA, that of laser energy absorption and electron heating [21]. The laser, target and simulation parameters are listed in Table 1. A sketch of the computational domain and target is shown in Figure 1. The laser pulse intensity has the form $I(t, y) = I_0 \sin^2(\pi t / 2\tau_{FWHM}) \exp(-(y/r_0)^2)$, where $r_0 = \frac{1}{2\sqrt{ln(2)}} D_{FWHM}$ is the focal spot radius at $1/e$ level. The parameters $D_{FWHM}$ and $\tau_{FWHM}$ are the laser spot size and pulse duration, respectively. The laser energy is calculated according to $\varepsilon_{laser} = \pi r_0^2 I_0 \tau_{FWHM} \cong 1.13 D_{FWHM}^2 I_0 \tau_{FWHM}$. The laser pulse propagates in the "+x" direction and is linearly polarized in the "y" direction. Particles are initialized with charge +1 for ions and −1 for electrons. During the simulations the ion charge of oxygen and gold is dynamically incremented using a standard Monte Carlo scheme [28,29]. Thus every computational particle has its own charge, which except for electrons and protons, may change in time.

Before presenting simulation results, it is worth mentioning the main characteristics of RPA and BoA. The fundamental difference between BoA and the linear RPA is the plasma transparency to the electromagnetic radiation. In the linear RPA regime, the plasma remains overdense and the laser pushes a double-layer structure of electrons and ions in a piston-like fashion. The electrons and ions form a co-moving sheath, which allows ions to be accelerated. In contrast, in BoA the main acceleration stage begins after the target has expanded and the electron density has dropped sufficiently so that the plasma becomes relativistically transparent to the laser pulse ($n_e < \gamma n_{cr}$), thus allowing the laser pulse to punch through the target. Because the target is transparent to the laser light, the electrons in and around the focal spot can regain their energy lost for accelerating ions, thus continuously accelerating the ions. Since initially the target is highly overdense, $n_e / n_{cr} > 1000$, transparency is achieved by plasma expansion in which the electron density decreases. But for BoA to work, the plasma must become transparent near the peak of the laser pulse. Since this process takes time, it is more typical for longer pulses, on the order of 0.5-1 ps. For very short pulses, 30-50 fs, the plasma remains overdense for the duration of the laser pulse and if the foil is thin enough, the main ion acceleration mechanism is RPA. Since plasma transparency is the key, in what follows we will assess it in order to attribute the acceleration to either BoA or RPA.

The presence of plasma transparency is usually established by plotting the electron density and comparing it to the relativistically corrected critical density. In 2D and 3D simulations complications arise since the electron density can vary widely from one location to another making assessment accurate locally, but somewhat difficult globally. We use a slightly different approach in which judgement is based on the laser pulse characteristics, specifically electromagnetic wave transmission and reflectivity. Though information regarding the plasma transparency is retrieved indirectly and may be subject to interpretation, the method is rather simple and can be readily applied. It is based on the global (integrated over the computational domain) energy balance, which at any given time reads [26,30]:

$$\varepsilon^{in}(t) = \varepsilon^{field}(t) + \varepsilon^{out}(t) + \varepsilon^{kin}(t). \tag{1}$$

Equation (1) expresses the fact that the electromagnetic wave energy which entered the computational domain prior to time $t$, $\varepsilon^{in}$, is balanced by the electromagnetic field energy $\varepsilon^{field}$ residing in the computational domain, electromagnetic energy $\varepsilon^{out}$ that has left in the computational domain, and specie kinetic energy $\varepsilon^{kin}$, the latter being summed over the kinetic energies of all computational particles. Figure 2 shows the individual components for three different cases, very short pulse (32 fs), intermediate case (180 fs) and long pulse (600 fs), whose characteristics are listed in Table 1. Time $t_0 = -160\ fs$ corresponds to the moment the laser pulse enters the computational domain at spatial position $x = 0$, and time $t = 0$ is the moment it reaches the target. For time $t \leq 0$ $\varepsilon^{field}(t) = \varepsilon^{in}(t)$, i.e. the energy entering the computational domain stays as energy of the electromagnetic field since there is no interaction with the target. At time $t = 0$ the laser bullet reaches the foil. Shortly thereafter, within 1-2 laser cycles, a hot and highly overdense plasma is formed within the target, which gradually increases to density in excess of $10^3$ times the critical electron density. Part of the electromagnetic pulse is reflected from the plasma mirror and turns around, while the transmitted part couples energy to the plasma.

In Figure 2a, one can observe a sharp increase of $\varepsilon^{out}$ starting at around $t = 160\ fs$. This is the reflected pulse going in the $-x$ direction, which has reached the computational domain edge ($x = 0$) and leaves. The peak of $\varepsilon^{out}$ can be used to estimate the reflection coefficient of the plasma, $\xi^r = \varepsilon^{out}(t_{sims})/\varepsilon^{laser}$. The simulations show that $\xi^r \cong 80\%$ of the laser energy is reflected by the plasma mirror, a strong indication that the plasma remains overdense. For the intermediate pulse length the electromagnetic pulse is partially reflected and transmitted, while for the long pulse the reflection is negligible. The largest fraction of the initial energy is retained as energy of the electromagnetic field (Figure 2c).

The coupling of laser energy to the plasma is also very informative. As seen in Figure 2a, for $t > 0$ $\varepsilon^{field}$ starts to decrease, while $\varepsilon^{kin}$ starts to increase. Particles gain energy from the electromagnetic field and the process is irreversible. However, for the long pulse (Figure 2c), both the electromagnetic field energy and kinetic energy increase with time, indicating that electromagnetic energy flows into the system and then continuously transferred to the plasma. This is one of the main characteristics of the BoA mechanism. Considering both the reflection of laser pulse and the traits of energy coupling, we conclude that for the short pulse illustrated in Figure 2a the ion acceleration mechanism is RPA, while for the long pulse (Figure 2c) it is BoA. The intermediate case is somewhat closer to RPA, but the analysis is inconclusive.

## 3. Simulations results for the Texas Petawatt Laser

Selected results for the Texas Petawatt Laser are shown in Figures 3-5. The momentum distribution of gold ions and spectrum in the forward direction at the end of the simulations (540 fs) are shown in Figure 3. Only energetic ions with kinetic energy >100 MeV (>0.5 MeV/nucleon) are included. About half of the ions are accelerated in the forward direction and the other half is scattered in the backward direction, toward the laser. The latter is most likely due to Coulomb explosion of unbalanced charge. The maximum normalized ion momentum in the forward direction is $p_x/Mc \cong 0.15$, which is less than that typically observed for protons and light ions.

The ion beam is characterized with its flux, spectrum, charge distribution, and angular distribution. For all of them we consider only energetic ions (>0.5 MeV/nucleon). In addition, except for the angular distribution, only ions lying within 10 degrees half-angle from the target normal are collected, which corresponds to solid angle $d\Omega = 0.095\, sr$. The spectrum of gold ions, $\frac{d^2N}{dEd\Omega}$, is plotted in Figure 3b. The energy distribution decreases exponentially and has a maximum energy of ~2 GeV (10 MeV/nucleon). The flux in the forward direction is $\frac{dN}{d\Omega} \cong 8\times 10^{11}\, \frac{ions}{sr}$. The gold ions charge distribution in the forward direction is shown in Figure 4a. The average charge-to-mass ratio is rather low, $\bar{q}/M \approx 0.2$. The angular distribution is highly peaked, which leads to a large flux in the forward direction. Most ions lie in a cone of ~20 degrees from the target normal (Figure 4b). There is a group of ions scattered backward, seen also in the phase-space plot in Figure 3a, presumably from Coulomb explosion of the Au layer. According to the simplified theory of RPA, the ions located initially in the compression layer will undergo RPA and will be snow-plowed forward because for these ions the electrostatic pressure balances the radiation pressure, while the plasma containing a sheath of bare ions in the electron depletion layer will Coulomb explode launching ions in the backward direction [15]. It is interesting to note that both forward accelerated and backward scattered ions have very narrow angular distributions, i.e. both are emitted perpendicular to the foil surface.

Complementary results are given for protons emitted from the contaminant layer on the back surface of the foil. The proton momentum distribution and spectrum in the forward direction are plotted in Figure 5. Only protons with energy >1 MeV are included. The proton spectrum shows a peak at ~10 MeV and the maximum proton energy is ~60 MeV. Comparing the maximum energy per nucleon of gold ions and protons, 10 and 60 MeV/nucleon, respectively, we observe a nearly linear dependence between the energy per nucleon and the charge-to-mass ratio:

$$\left(\frac{E}{M}\right)_{max} \sim \left(\frac{q}{M}\right)_{max} \qquad (2)$$

Our findings are in contrast to previous studies, which showed a much stronger (quadratic)

dependence.

## 4. Simulations results for the Trident Laser

The Trident laser has been extensively used for ion acceleration in the BoA regime. The focus was on light ions: protons [31], deuterons [31,32] and carbon ions [20,21,22,33,], but ions from mid-Z material (Pd) were accelerated as well [34,35]. The results are impressive: deuterons with normalized energy of 40 MeV/nucleon and $C_6^+$ ions with energies exceeding 80 MeV/nucleon have been measured. In all cases the major role played the long pulse, which ensured long continuous acceleration of electrons and ions. The same reasoning should apply to heavy ions. Simulation results for the Trident laser are shown in Figures 6-8. The momentum distribution and spectrum of gold ions are similar to that of the Texas Petawatt laser, but the ion energies are higher, presumably due to the longer acceleration time. Gold ions from the bulk of the foils are also blown in the backward direction. This feature of heavy ion acceleration was observed in all simulations. The charge distribution is, however, different. Almost 1/2 of the forward accelerated gold ions have charge $q = 51$, and the maximum ion charge is 62, which contributes to the observed higher maximum ion energy (Figure 7). The gold ion angular distribution, both in the forward and backward directions, is broader. We surmise that this is a result of Coulomb explosion of the forward accelerated beam of ions. The effect is more pronounced for the Trident laser for two reasons: the ion charge in the beam is three times larger (see Table 2) and the snapshot is done at a later time (1.2 ps vs. 0.54 ps), and the focal spot size of the Trident laser is twice smaller compared to the Texas Petawatt Laser (5 μm versus 10 μm).

The most profound difference between the two laser systems is found for the proton beam (Figure 8). The ion spectrum contains the expected exponential distribution, but in addition, it has a well-defined monoenergetic component centered around 100 MeV. The total number of monoenergetic protons is ~$10^{11}$.

## 5. Discussion

A summary of simulation results for the ion beam properties for the three laser systems is given in Table 2, where the calculated ion fluxes and maximum energy per nucleon in the forward direction are listed. For gold, the ion fluxes are comparable, on the order of $10^{11}$ ion/sr, and not directly related to the laser energy on target. The highest flux is for the laser with intermediate energy (Texas Petawatt Laser) due to its largest focal spot size (compare in Table 1), while the highest energy per nucleon is for the laser with the longest pulse duration (Trident). The same holds for protons, except the proton flux (from the contaminant layer) is one order of magnitude larger than the flux of gold ions (from the bulk of the foil). Both protons and gold ions are in quantities that makes them detectable with a Thompson parabola. For a typical opening angle $\Delta\Omega \approx 10^{-7}\, sr$, we estimate that about $10^5$ protons and $10^4$ gold ions are expected to enter the Thompson parabola.

The numerical simulations indicate that the ion acceleration mechanisms for the three laser systems are very different: RPA for the shortest pulse, BoA for the longest and somewhat undetermined in the intermediate case. But regardless of the vast differences in the underlying physical mechanisms of the acceleration process, the results, e.g. ion beam parameters, are not dissimilar. All laser systems considered are suitable for heavy ion acceleration, which is mostly due to the large amount of energy (>20 J) delivered on target.

## 6. Conclusion

In conclusion, acceleration of heavy ions from ultrathin (<<1 μm) foil has been investigated theoretically using a 2D PIC code. The theory of ion acceleration mechanisms developed for light ions was extended to heavy ions. Nuances of the acceleration process, in particular, the two stages of ion acceleration (ionization phase followed by acceleration phase) differentiates the heavy ion acceleration. Another distinct feature is the much more demanding requirements about the laser system, which are needed in order to overcome difficulties inherent for heavy ion acceleration. We demonstrated numerically that laser systems capable of delivering sufficient amount of energy on target (>10 Joules) can generate ion fluxes with sufficient energy and number. Numerical simulations show that three such systems, Bella, Texas Petawatt Laser and Trident can generate a beam of gold ions with energy 10-45 MeV/nucleon, flux $10^{11}$ ion/sr, and angular divergence of 20 degrees.

**Acknowledgements:**

This work was performed with the support of the Air Force Office of Scientific Research under grant FA9550-14-1-0282. G. M. P. would like to acknowledge the DoD HPC computing program at NRL.

**Figure captions:**

Figure 2. Energy balance components in Equation 1 versus time for the Bella laser (a), Texas Petawatt Laser (b) and Trident laser system (c): energy entering the computational domain $\varepsilon^{in}$, energy leaving the computational domain $\varepsilon^{out}$, electromagnetic field energy $\varepsilon^{field}$ and kinetic energy $\varepsilon^{kin}$. Time $t_0 = -160 \, fs$ corresponds to the moment the laser pulse enters the computational domain and time $t = 0 \, fs$ is the moment the laser pulse reaches the target. The laser and foil parameters are listed in Table 1.

Figure 3. Momentum distribution of gold ions (a) and energy spectrum of forward accelerated gold ions (b) at the end of the simulations for the Texas Petawatt Laser. Only ions with energy >100 MeV within solid angle $d\Omega = 0.095 \, sr$ are included. The total ions flux is $\frac{dN}{d\Omega} \cong 7.7 \times 10^{11} \frac{ions}{sr}$.

Figure 4. Charge distribution of forward accelerated gold ions (a) and angular distribution of gold ions (b) at the end of the simulations for the Texas Petawatt Laser.

Figure 5. Momentum distribution of protons (a) and energy spectrum of forward accelerated protons (b) at the end of the simulations for the Texas Petawatt Laser. Only ions with energy >1 MeV within solid angle $d\Omega = 0.095 \, sr$ are included. The total protons flux is $\frac{dN}{d\Omega} \cong 4.0 \times 10^{12} \frac{ions}{sr}$.

Figure 6. Momentum distribution of gold ions (a) and energy spectrum of forward accelerated gold ions (b) at the end of the simulations for the Trident laser. Only ions with energy >100 MeV within solid angle $d\Omega = 0.095 \, sr$ are included. The total ions flux is $\frac{dN}{d\Omega} \cong 2.6 \times 10^{11} \frac{ions}{sr}$.

Figure 7. Charge distribution of forward accelerated gold ions (a) and angular distribution of gold ions (b) at the end of the simulations for the Trident laser.

Figure 8. Momentum distribution of protons (a) and energy spectrum of forward accelerated protons (b) at the end of the simulations for the Trident laser. Only ions with energy >1 MeV within solid angle $d\Omega = 0.095 \, sr$ are included. The total protons flux is $\frac{dN}{d\Omega} \cong 1.2 \times 10^{12} \frac{ions}{sr}$.

Table 1. Laser, target and computational domain parameters used in the simulations.

| parameter | variable & units | Bella | TPW | Trident |
|---|---|---|---|---|
| laser intensity | $I_0$ (W/cm$^2$) | $3\times10^{21}$ | $2.5\times10^{20}$ | $5\times10^{20}$ |
| pulse duration | $\tau_{FWHM}$ ($\mu m$) | 32 | 180 | 600 |
| focal spot size | $D_{FWHM}$ ($\mu m$) | 5 | 10 | 5 |
| wavelength | $\lambda$ ($\mu m$) | 0.8 | 1 | 1 |
| energy | $\varepsilon_{laser}$ (J) | 27 | 50 | 80 |
| foil thickness | $L$ (nm) | 20 | 20 | 20 |
| foil width | $W$ ($\mu m$) | 126 | 126 | 254 |
| foil location | $x_0$ ($\mu m$) | 48 | 48 | 48 |
| computational domain | $L_x \times L_y$ ($\mu m^2$) | 100x128 | 100x128 | 200x256 |
| cell size | $\Delta x \times \Delta y$ (nm$^2$) | 20x20 | 20x20 | 20x20 |
| time step | $\Delta t$ ($\lambda/c$) | 0.005 | 0.005 | 0.005 |
| simulation time | $t_{sims}$ (fs) | 320 | 540 | 1200 |

Table 2. Calculated flux and maximum energy per nucleon in the forward direction for protons and gold ions. The laser and foil parameters are listed in Table 1.

| parameter | variable & units | Bella | TPW | Trident |
|---|---|---|---|---|
| pulse duration | $\tau_{FWHM}$ (fs) | 32 | 180 | 600 |
| energy | $\varepsilon_{laser}$ (J) | 27 | 50 | 80 |
| Au ions | $dN/d\Omega$ (ion/sr) <br> $(E/M)_{max}$ (MeV/nucleon) | $1.7\times10^{11}$ <br> 25 | $7.7\times10^{11}$ <br> 10 | $2.6\times10^{11}$ <br> 45 |
| protons | $dN/d\Omega$ (ion/sr) <br> $(E/M)_{max}$ (MeV/nucleon) | $2.2\times10^{12}$ <br> 85 | $4.0\times10^{12}$ <br> 60 | $1.2\times10^{12}$ <br> 120 |

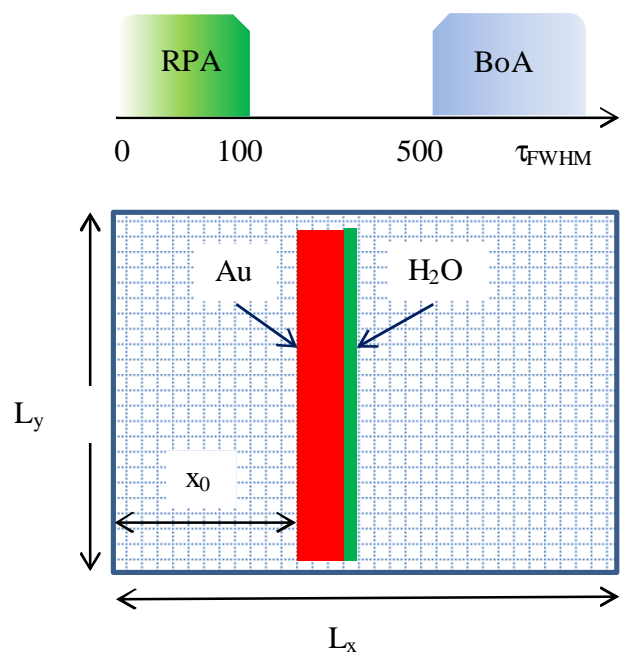

Figure 1

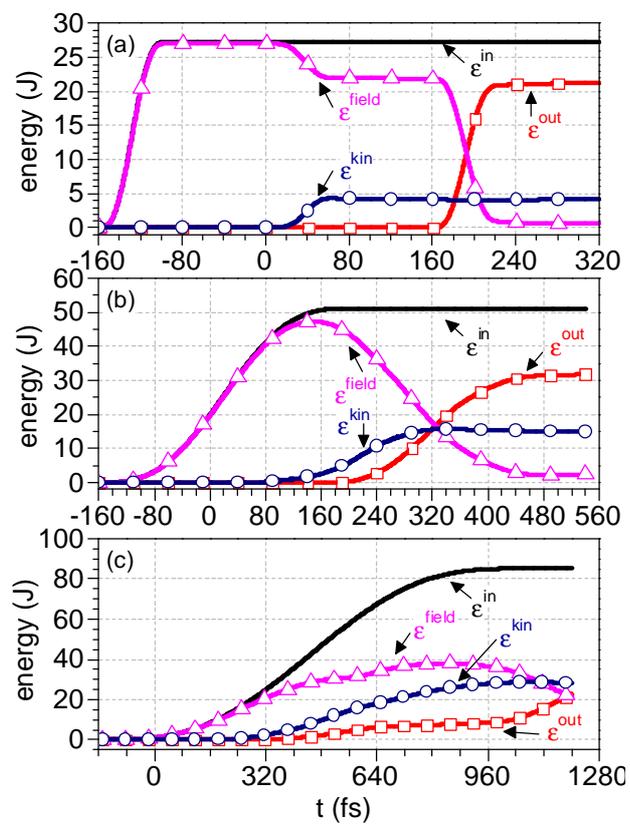

Figure 2

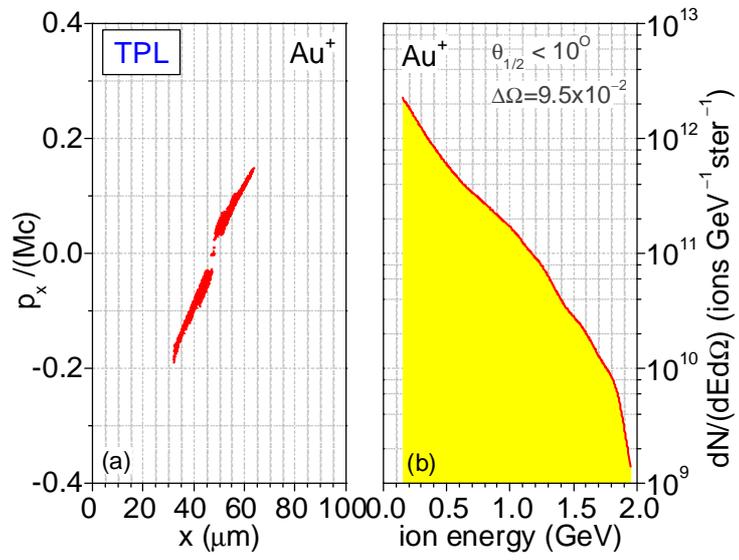

Figure 3

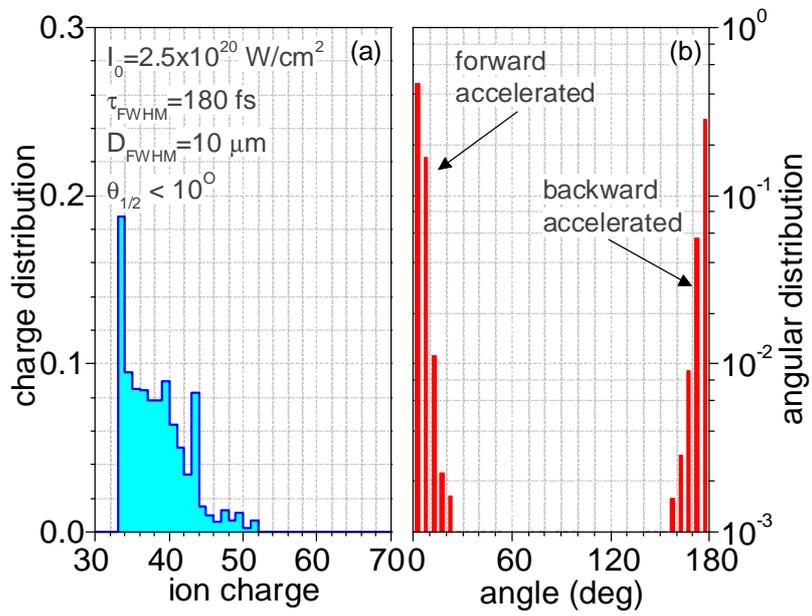

Figure 4

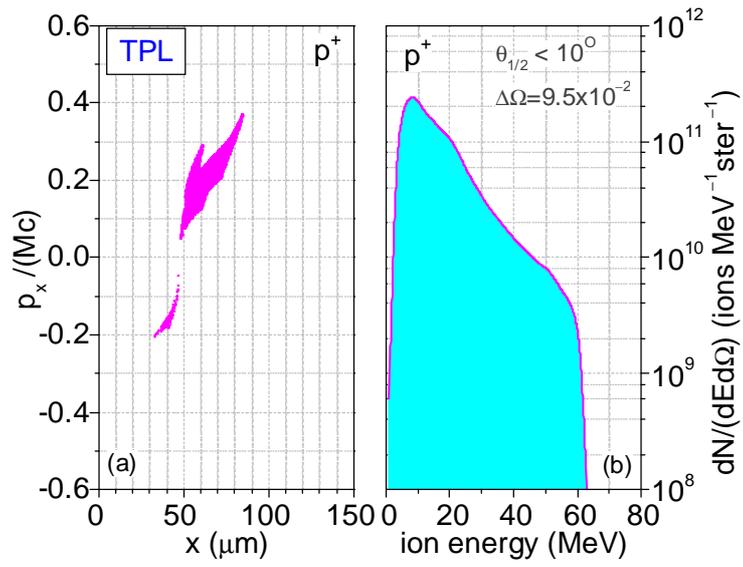

Figure 5

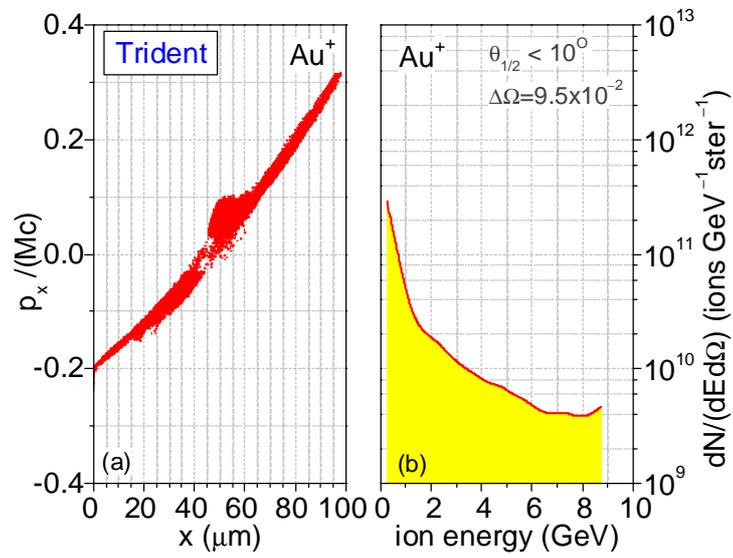

Figure 6

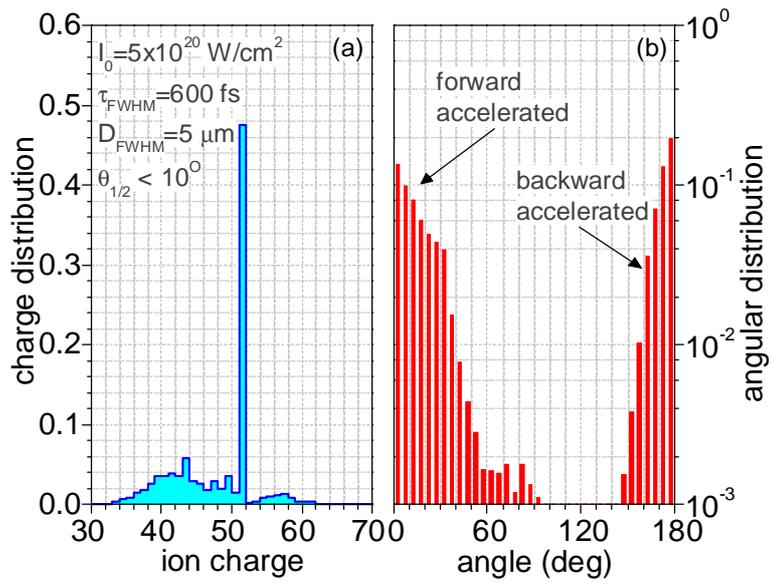

Figure 7

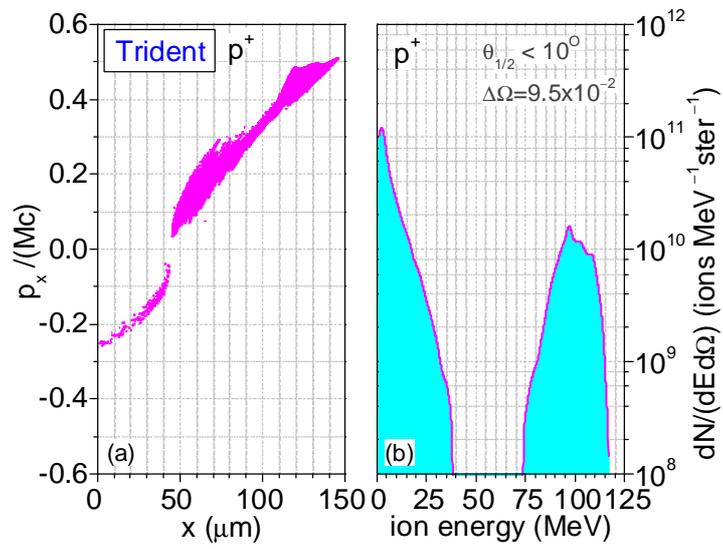

Figure 8